\documentclass[%
 aip,
 amsmath,amssymb,
reprint,%
]{revtex4-2}

\usepackage{graphicx}% Include figure files
\usepackage{dcolumn}% Align table columns on decimal point
\usepackage{upgreek}
\usepackage{bm}

\usepackage[utf8]{inputenc}
\usepackage[T1]{fontenc}
\usepackage{mathptmx}
\usepackage{etoolbox}
\usepackage{diagbox}

\usepackage{amsmath}
\usepackage{color}
\usepackage{xfrac}

\usepackage{caption}
\usepackage{subcaption}

\usepackage{graphicx}
\usepackage{floatrow}
\graphicspath{{figures/}}

\usepackage{amssymb}
\usepackage{amsbsy}
\usepackage{textcomp}
\usepackage{natbib}

\newcommand{\RomanNumeralCaps}[1]
\linenumbers

\makeatletter
\def\@email#1#2{%
 \endgroup
 \patchcmd{\titleblock@produce}
  {\frontmatter@RRAPformat}
  {\frontmatter@RRAPformat{\produce@RRAP{*#1\href{mailto:#2}{#2}}}\frontmatter@RRAPformat}
  {}{}
}%
\makeatother
\begin{document}

\preprint{AIP/123-QED}

\title{Three-dimensional computational rheometry of viscoplastic fluids in vane-in-cup fixtures}
% Force line breaks with \\

% altaffiliation
\author{James Malcolm}
\author{Clare McColgan}
\author{Emma Mullarkey}
\author{Emad Chaparian}
 \email{emad.chaparian@strath.ac.uk}
\affiliation{James Weir Fluid Laboratory, Department of Mechanical \& Aerospace Engineering, University of Strathclyde, Glasgow, United Kingdom, G1 1XJ}

\date{\today}% It is always \today, today,

\begin{abstract}
We develop a computational framework to investigate the flow kinematics of viscoplastic materials, described by the Herschel-Bulkley model, in vane-in-cup rheometers. Numerical simulations are performed using augmented Lagrangian finite element and regularised finite volume methods for two- and three-dimensional configurations, respectively, to characterise the hydrodynamics and yielding behaviour over a broad range of effective parameters. Three types of vanes are investigated: conventional 4- and 6-arm bespoke vanes and a 24-arm fractal vane (Owens {\it et al.}, J.~Rheol., vol.~64(3), 2020, 643-662). Our previous two-dimensional study (Chaparian {\it et al.}, J.~Non-Newtonian Fluid Mech., vol.~307, 2022, 104857) demonstrated the superiority of the fractal vane over conventional $N$-arm designs. The bespoke vanes produce highly non-uniform yielding, leading to significant deviations from the circular Couette analogy, whereas the 24-arm fractal vane generates nearly axisymmetric yield surfaces that effectively ``cloak'' the vane geometry. Consequently, the resulting flow closely resembles the azimuthal shear in a conventional ``bob \& cup'' rheometer, providing substantially more accurate rheological measurements. Building on these findings, the present study revisits the torque-to-stress and rotation-to-strain rate conversions used in vane-in-cup rheometry. In particular, we re-examine the circular Couette analogy that underpins both the physical interpretation of the flow and the extraction of flow curves from raw rheometric data. Based on our extensive two- and three-dimensional simulations, we identify limitations of the existing conversion relations and propose improved closure expressions that enable more accurate determination of material's rheological properties. Furthermore, the three-dimensional flow structures and yielding patterns within the vane-in-cup fixtures are characterised.
\end{abstract}

\maketitle
 
\section{Introduction}\label{sec:intro}

\begin{figure}
\includegraphics[width=\linewidth]{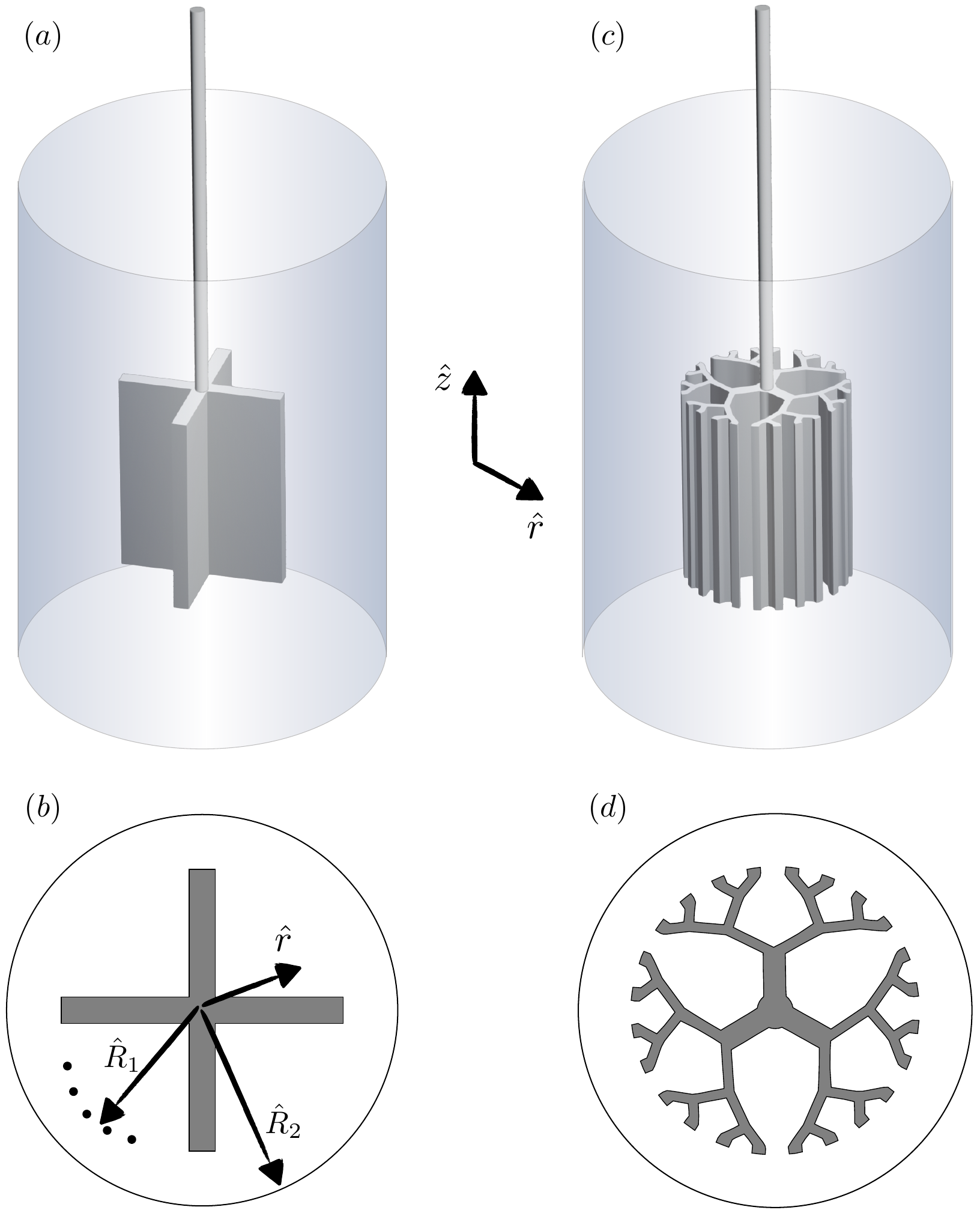}
\caption{Schematic of the vane-in-cup fixtures and the cross section of the 4-arm bespoke and the 24-arm fractal vanes.}
\label{fig:schematic}
\end{figure}

``Bob \& cup'' rheometers are widely used for characterising the rheological behaviour of simple liquids, where the sample is subjected to azimuthal shear within the gap between the bob and the cup. However, their application is not always feasible for complex industrial materials. For instance, many practical materials contain large particles, such as fibre suspensions \cite{derakhshandeh2010rheology} and cement slurries. Moreover, many of these materials exhibit ``apparent slip'' \cite{meeker2004PRL,daneshi2019characterising,chaparian2021sliding} on smooth surfaces, which introduces a barrier in using a perfect cylindrical bob as the driving element for shear. As an alternative, vane-in-cup fixtures (see Fig.~\ref{fig:schematic}(a,b)) have become a widely used rheological measurement tool for such materials \cite{barnes2001rotating,ovarlez2011flows}, with applications ranging from cement slurries and concrete characterisation \cite{ferraris2001comparison,ferraris2004comparison} to food processing \cite{steffe1996rheological} and bioengineering/pharmaceutical products \cite{barnes2001rotating}. Many of these materials are classified as yield-stress fluids: they exhibit a ``solid-like'' response when the applied stress is below a critical threshold (the yield stress, $\hat{\tau}_y$) and flow once this threshold is exceeded. Such materials are widely encountered in everyday applications, including cosmetic foams, emulsions (e.g.~mayonnaise), microgels (e.g.~hand sanitisers and hair gels), edible products (e.g.~ketchup), etc. Nevertheless, vane-in-cup rheometry is not limited to yield-stress fluids; shear-thinning \& thixotropic \cite{barnes1990vane}, and viscoelastic \cite{vadillo2025vane,you2026identification} behaviours have also been characterised using these fixtures.

A major advantage of vane-in-cup fixtures, as mentioned, is their ability to reduce the influence of non-ideal slip, which can otherwise introduce significant errors in the measurement of rheological properties, particularly for yield-stress fluids. Despite this intuitive advantage, the extent to which vanes suppress slip has received relatively little attention. Recently, \citet{chaparian2022vane}, using a highly accurate computational framework for simulating ``stick-slip'' laws \cite{chaparian2021sliding}, demonstrated that an $N$-arm vane can effectively eliminate the entire effect of slip and produce rheological measurements identical to those obtained under no-slip conditions. In other words, conventional $N$-arm bespoke vanes are capable of entraining the viscoplastic material and generating the same level of shear as the corresponding no-slip case, even when the material tends to slide along the vane surfaces. Therefore, by combining such vanes with a roughened or grooved cup, or by applying other slip-mitigation techniques (e.g.~chemical treatment \cite{christel2012stick,jalaal2015slip,baxter2026rheology}), rheological measurements can be obtained without significant slip-induced errors. However, conventional $N$-arm bespoke vanes suffer from a fundamental limitation: they generate highly non-uniform shear fields. Using Magnetic Resonance Imaging (MRI), \citet{ovarlez2011flows} showed that the yield surface surrounding a 6-arm vane deviates substantially from a circular shape, resulting in a yielded region whose thickness varies with the azimuthal angle. \citet{chaparian2022vane} further investigated this issue and showed that, although increasing the number of arms improves the shear uniformity, even $N=12$ remains insufficient: the extensional deformation rate experienced by fluid elements near the vane tips can reach approximately $ 10\% $ of the shear rate.

To address this limitation, Owens {\it et al.}~\cite{owens2020improved,FractalVanePatent} introduced a class of modified vanes with fractal cross-sectional geometries; see Fig.~\ref{fig:schematic}(c,d). These designs significantly improve the azimuthal kinematics uniformity, producing smoother yield surfaces near the vane tips and a circular outer yield surface. Using two-dimensional numerical simulations over a wide range of effective parameters, \citet{chaparian2022vane} investigated the azimuthal flow kinematics generated by fractal vanes and compared their performance with conventional $N$-arm vane geometries. Subsequently, \citet{medina2023rheo} experimentally assessed the performance of fractal vanes using three representative fluids: glycerol (a purely viscous fluid), Carbopol gel (a ``simple'' yield-stress fluid), and $\kappa$-Carrageenan gel (a thixotropic yield-stress fluid). Using Particle Image Velocimetry (PIV), they measured the velocity field of the sheared sample in the radial gap. For glycerol and Carbopol gel, the measurements obtained using the 24-arm fractal vane were consistent with conventional rheometric techniques, and the PIV measurements showed good agreement with numerical predictions. For $\kappa$-Carrageenan gel, however, shear banding was observed, unlike the behaviour observed for a conventional 6-arm vane. This was attributed to disruption of the material microstructure during immersion of the 24-arm vane, which triggered syneresis. Very recently, \citet{vadillo2025vane} extended the investigation of fractal vanes to viscoelastic fluids and provided further analysis of shear-banding phenomena.

One of the crucial challenges in vane-in-cup rheometry is the accurate conversion of the measured torque and vane rotation rate into the nominal stress and strain rate through appropriate closure relations \cite{Macosko1994_TCintro,Macosko1994_endeffects}. Such conversions are essential for making vane-in-cup measurements comparable with conventional shear rheometry techniques, including plate \& plate and cone \& plate geometries. Accurate closure relations must account for two key effects: (i) the inherent inhomogeneity of the shear field generated by the vane geometry, and (ii) three-dimensional flow effects \cite{nguyen1987characterization,sherwood1991use,atkinson1992torque}. Although several modified bob geometries (e.g.~the Mooney-Ewart geometry) have been proposed to reduce three-dimensional effects, these effects cannot be completely eliminated. In vane-in-cup rheometry, where the flow field is intrinsically more complex, the development of accurate conversion relations is therefore particularly important.

In the present study, we computationally investigate the three-dimensional flow kinematics of a viscoplastic material in vane-in-cup fixtures with conventional 4- and 6-arm bespoke vanes and a 24-arm fractal vane. One of the primary objectives is to develop a robust numerical framework for three-dimensional simulations of flows within these type of rheometers. Furthermore, by combining extensive two- and three-dimensional simulations, we assess the validity of existing stress and strain rate conversion relations and propose improved formulae for extracting flow curves from raw vane-in-cup rheometry data.

The remainder of the paper is organised as follows. Section~\ref{sec:formulation} introduces the problem formulation, followed by the numerical methodology in \S \ref{sec:method}. The flow-curve calculations based on the two-dimensional simulations are presented in \S \ref{sec:2D}. Section~\ref{sec:3D} examines the influence of three-dimensional effects on the flow-curve calculations and discusses the flow kinematics associated with each vane geometry. Finally, the main conclusions are summarised in \S \ref{sec:conclusion}.

\section{Problem formulation}\label{sec:formulation}

We consider an inertialess incompressible isothermal steady-state flow of a ``simple'' yield-stress fluid in a vane-in-cup fixture. Hence, the Cauchy and the continuity equations governing this problem are:
\begin{eqnarray}\label{eq:dimNS}
	0 &=& - \boldsymbol{\nabla} \hat{p} + \boldsymbol{\nabla} \boldsymbol{\cdot} \hat{\boldsymbol{\uptau}}, ~~~\text{in}~~ \Gamma \setminus \bar{X}, \\
0 &=& \boldsymbol{\nabla} \boldsymbol{\cdot} \hat{\boldsymbol{u}} , ~~~\text{in}~~ \Gamma \setminus \bar{X}.
\end{eqnarray}
where $\hat{\boldsymbol{u}}=(\hat{u}_r,\hat{u}_{\theta},\hat{u}_z)$ is the velocity vector, $\hat{p}$ is the pressure, and $\hat{\boldsymbol{\uptau}}$ is the deviatoric stress tensor, while the Cauchy stress (or the total stress) tensor is $ \hat{\boldsymbol{\sigma}} = - \hat{p} \boldsymbol{1} + \hat{\boldsymbol{\uptau}} $, where $\boldsymbol{1}$ is the identity matrix. The full domain (i.e.~cup space) is $\Gamma$ and the vane occupies $X$. The fluid's rheological behaviour follows the Herschel-Bulkley model,
\begin{equation}\label{eq:const}
	\left\{
	\begin{array}{ll}
		\hat{\boldsymbol{\uptau}} = \left( \hat{K} \Vert \hat{\dot{\boldsymbol{\upgamma}}} \Vert ^{n-1} + \displaystyle{\frac{\hat{\tau}_y}{\Vert \hat{\dot{\boldsymbol{\upgamma}}} \Vert}} \right) \hat{\dot{\boldsymbol{\upgamma}}} & \mbox{iff}\quad \Vert \hat{\boldsymbol{\uptau}} \Vert > \hat{\tau}_y, \\[2pt]
		\hat{\dot{\boldsymbol{\upgamma}}} = 0 & \mbox{iff}\quad \Vert \hat{\boldsymbol{\uptau}} \Vert \leqslant \hat{\tau}_y,
	\end{array} \right.
\end{equation}
where $\hat{\dot{\boldsymbol{\upgamma}}}$ is the  rate-of-strain tensor. Rheological parameters are $\hat{\tau}_y$, the yield stress of the material, $\hat{K}$, the consistency, and the power-law index $n$. The von Mises yielding criterion is considered for the material with the yield stress $\hat{\tau}_y$, i.e.~$\Vert \cdot \Vert$ is the second invariant of a tensor or $\Vert \boldsymbol{\Lambda} \Vert = \sqrt{(1/2) \boldsymbol{\Lambda} \boldsymbol{:} \boldsymbol{\Lambda}}$ when $\boldsymbol{\Lambda}$ is deviatoric.

\begin{figure*}
\centerline{\includegraphics[width=0.9\linewidth]{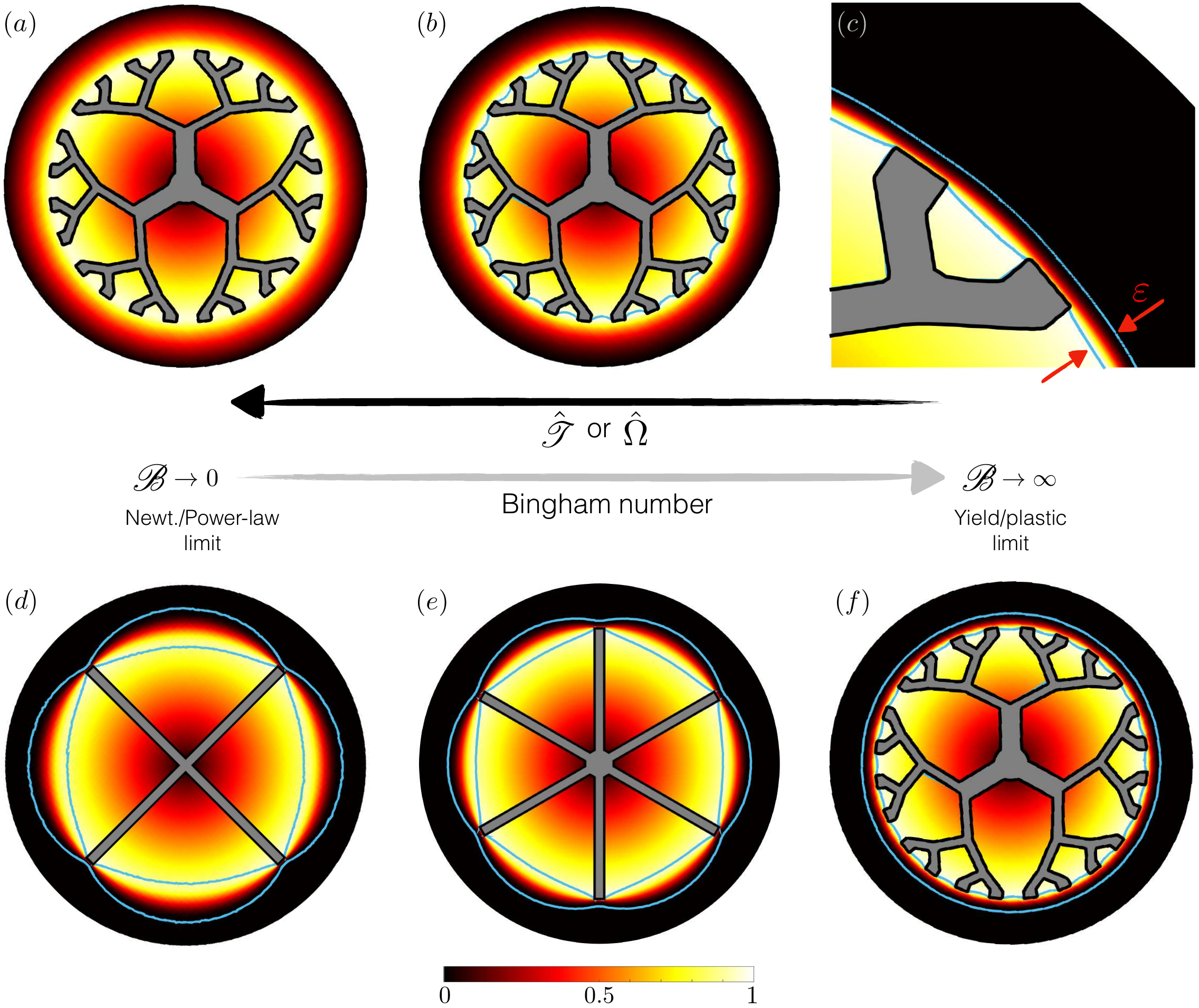}}
\caption{Interpretation of the Bingham number in terms of the applied torque/rotation rate on the vane. The contours show the dimensionless velocity magnitude $\vert \boldsymbol{u} \vert$ of the 2D computations. Panels (a-c) represent the case of a 24-arm fractal vane, from left to right at $\mathcal{B}=0$, $\mathcal{B}=15.43$, and $\mathcal{B}=100$, respectively. Panel (c) is rather an enlarged view focusing on the thin yielded boundary layer with thickness $\epsilon$. The blue lines denotes the yield surfaces, i.e.~$\Vert \tau \Vert = \mathcal{B}$ lines. Panels (d-f) show the flow fields at $\mathcal{B}=15.43$; from left to right, the 4- and 6-arm bespoke and 24-arm fractal vanes, respectively. All the panels are associated with the case $\mathbb{A}$ geometry and the fluid power-law index $n=0.46$.}
\label{fig:Torque_Bingham_axis}
\end{figure*}

We scale the above equations with the linear velocity of the vane tip (i.e.~$\hat{R}_1 \hat{\Omega}$) and using the gap width (i.e.~$\hat{\ell}=\hat{R}_2 - \hat{R}_1$) as the length scale where $\hat{R}_1$ is the vane's arm length and $\hat{R}_2$ is the cup radius (see Fig.~\ref{fig:schematic}(b)) and $\hat{\Omega}$ is the rotational rate of the vane. Also, the deviatoric stress tensor and the pressure are scaled with the characteristic viscous stress, hence,
\[
\hat{\boldsymbol{u}} = \hat{R}_1 \hat{\Omega} ~\boldsymbol{u}, ~~\hat{\boldsymbol{\dot{\upgamma}}} = \frac{\hat{R}_1 \hat{\Omega}}{\hat{\ell}} \boldsymbol{\dot{\upgamma}}, ~~(\hat{\boldsymbol{\uptau}},\hat{p}) = \hat{K} \left( \frac{\hat{R}_1 \hat{\Omega}}{\hat{\ell}} \right)^n (\boldsymbol{\uptau},p), ~~
\]
yielding dimensionless equations,
\begin{equation}\label{eq:non-govern}
0 = - \boldsymbol{\nabla} p + \boldsymbol{\nabla} \boldsymbol{\cdot} \boldsymbol{\uptau}~~~\text{in}~~ \Gamma \setminus\bar{X},
\end{equation}
and,
\begin{equation}\label{eq:non-const}
  \left\{
    \begin{array}{ll}
      \boldsymbol{\uptau} = \left( \Vert \dot{\boldsymbol{\upgamma}} \Vert^{n-1} + \displaystyle{\frac{\mathcal{B}}{\Vert \dot{\boldsymbol{\upgamma}} \Vert}} \right) \dot{\boldsymbol{\upgamma}} & \mbox{iff}\quad \Vert \boldsymbol{\uptau} \Vert > \mathcal{B}, \\[2pt]
      \dot{\boldsymbol{\upgamma}} = 0 & \mbox{iff}\quad \Vert \boldsymbol{\uptau} \Vert \leqslant \mathcal{B},
  \end{array} \right.
\end{equation}
where $\mathcal{B}$ is the Bingham number, 
\[
\mathcal{B} = \frac{\hat{\tau}_y}{\hat{K}} \left( \frac{\hat{\ell}}{\hat{R}_1 \hat{\Omega}} \right)^n,
\]
showing the ratio of the yield stress to the characteristic viscous stress. In this scaling, $\mathcal{B} \to \infty$ corresponds to the yield limit where the applied torque on the vane is just sufficient to overcome the material's yield resistance and initiate the rotation. On the other side of the spectrum, $\mathcal{B} \to 0$ represents the regime in which the yield stress of the material is negligible compared to the applied stress on the sample (or equivalently, the applied torque on the vane), reducing the problem to the Newtonian or power-law limit, depending on the value of $n$. This is illustrated in Fig.~\ref{fig:Torque_Bingham_axis}.

The torque $\hat{\mathcal{T}}$ can be computed directly by integrating the couple resulting from the stress acting on the vane surface,
\begin{equation}\label{eq:torque}
\hat{\mathcal{T}} = \int_{\partial X} \hat{\boldsymbol{r}} \boldsymbol{\times} \left( \hat{\boldsymbol{\sigma}} \boldsymbol{\cdot} \boldsymbol{n} \right) ~\text{d}\hat{A},
\end{equation}
or alternatively from the energy balance equation, which, in a steady-state problem, reads,
\begin{equation}
\int_{\Gamma \setminus \bar{X}} \hat{\boldsymbol{\uptau}} \boldsymbol{:} \hat{\dot{\boldsymbol{\upgamma}}} ~\text{d}\hat{V} = \int_{\Gamma \setminus \bar{X}} \left[ \left( \hat{K} \Vert \hat{\dot{\boldsymbol{\upgamma}}} \Vert^{n-1} + \displaystyle{\frac{\hat{\tau}_y}{\Vert \hat{\dot{\boldsymbol{\upgamma}}} \Vert}} \right) \hat{\dot{\boldsymbol{\upgamma}}} \right] \boldsymbol{:} \hat{\dot{\boldsymbol{\upgamma}}} ~\text{d}\hat{V} =  \hat{\mathcal{T}} ~\hat{\Omega},
\end{equation}
and can be simplified as,
\begin{equation}\label{eq:energy}
\hat{K} \int_{\Gamma \setminus \bar{X}} \Vert \hat{\dot{\boldsymbol{\upgamma}}} \Vert^{n+1} ~\text{d}\hat{V} + \hat{\tau}_y \int_{\Gamma \setminus \bar{X}} \Vert \hat{\dot{\boldsymbol{\upgamma}}} \Vert  ~\text{d}\hat{V} = \hat{\mathcal{T}} ~\hat{\Omega}.
\end{equation}
In the above expression (\ref{eq:energy}), the first term in the left hand side is the viscous dissipation and the second term is the plastic dissipation. The right hand side is the work done by the vane which is the driver of the shear in this problem. The torque on the vane can be scaled as,
\begin{equation}\label{eq:torque_scaling}
\hat{\mathcal{T}} = \hat{K}\frac{(\hat{R}_1 \hat{\Omega})^n}{\hat{\ell}^{n-3}} \mathcal{T}.
\end{equation}
Hence, the non-dimensional form of expression (\ref{eq:energy}) is,
\begin{equation}\label{eq:energy_non}
\int_{\Gamma \setminus \bar{X}} \Vert \dot{\boldsymbol{\gamma}} \Vert^{n+1} ~\text{d}V + \mathcal{B} \int_{\Gamma \setminus \bar{X}} \Vert \dot{\boldsymbol{\gamma}} \Vert  ~\text{d}V = \mathcal{T} \Big( \frac{1 - R}{R}\Big),
\end{equation}
where $R=\hat{R}_1/\hat{R}_2$ is the ratio of the vane radius to the cup radius.

In what follows, the subscripts $_{2D}$ and $_{3D}$ are used to distinguish torques computed from 2D or 3D simulations, respectively. In 2D problems, the torque $\hat{\mathcal{T}}_{2D}$ has unit $N \cdot m/m$ since indeed it represents the torque per unit length of the vane. Therefore, if we are dealing with a 2D problem, the power of the length scale in the denominator of expression (\ref{eq:torque_scaling}) should be $n-2$.

\section{Computational details}\label{sec:method}

\subsection{Fixture configurations}
Regarding the fixture dimensions, we consider two configurations based on geometries previously used in experimental studies: (i) a narrow-gap configuration (case $\mathbb{A}$), with dimensions adopted from the 24-arm fractal vane fixture of \citet{medina2023rheo}, and (ii) a wide-gap configuration (case $\mathbb{B}$), with dimensions adopted from the 4-arm bespoke vane fixture of \citet{potanin20103d}. Note that all three vane geometries considered in the present study (4-arm bespoke, 6-arm bespoke, and 24-arm fractal) are investigated under both configurations, allowing the influence of the gap size to be assessed independently of the vane shape. The dimensions of the two configurations are summarised in Table~\ref{tab:dimensions}, where $\hat{L}$ denotes the vane length (in the $z$-direction) and the final column specifies the clearance between the bottom of the vane and the bottom of the cup. The corresponding radial gap widths between the vane tips and the cup wall are $\hat{\ell}=4.65~\mathrm{mm}$ and $\hat{\ell}=6.75~\mathrm{mm}$ for cases $\mathbb{A}$ and $\mathbb{B}$, respectively.

\begin{table}[h]
%\begin{ruledtabular}
\begin{tabular}{ c c c c c c c}
\hline
\hline
 &~~
\textrm{$\hat{R}_1$} ~&~
\textrm{$\hat{R}_2$} ~&~
\textrm{$R$}~&~
\textrm{$\hat{L}$}~&~
\begin{tabular}{@{}c@{}}Sample \\ height\end{tabular} ~&~
\begin{tabular}{@{}c@{}}Bottom \\ clearance\end{tabular}
\\
\colrule
Case $\mathbb{A}$  ~&~ $14$  ~&~ $18.65$  ~&~ $0.7507$  ~&~ $42$ & $50$ & $4$\\
\hline
Case $\mathbb{B}$  ~&~ $7$  ~&~ $13.75$  ~&~ $0.5091$  ~&~ $30$ & $54$ & $10$\\
\hline
\hline
\end{tabular}
%\end{ruledtabular}
\caption{\label{tab:dimensions}%
Vane-in-cup fixture dimensions (all values are in millimeter).}
\end{table}

In this study, no-slip boundary condition is enforced on the cup wall and the vane surface (i.e.~$\partial \Gamma$ and $\partial X$).

Two complementary numerical approaches are employed in this study: (i) anisotropic adaptive finite element approach based on the augmented Lagrangian method for 2D simulations, and (ii) regularised finite volume approach for 3D computations. The following subsections provide a detailed description of these methods.

\subsection{2D computational method}

The augmented Lagrangian method provides a robust and reliable framework for simulating yield-stress fluid flows \cite{glowinski2011numerical,roquet2003adaptive}. Over the past decades, it has been successfully used in a wide range of viscoplastic flow problems \cite{saramito2017progress}. In our previous works, we have employed this approach to study multiphase flows \cite{chaparian2017yield,chaparian2017cloaking,chaparian2018inline,chaparian2021clouds,pourzahedi2022flow,izadi2023,pourzahedi2024flow}, flows in porous media \cite{chaparian2019porous,fraggedakis2021first,chaparian2024yielding,chaparian2025darcy1,chaparian2025darcy2} and computational rheometry \cite{chaparian2022vane,medina2023rheo}. We have also extended this method to incorporate elastic effects \cite{chaparian2019adaptive} and ``apparent slip'' boundary conditions \cite{chaparian2021sliding,chaparian2022vane}.

To highlight, the augmented Lagrangian method is capable of handling the non-differentiable viscoplastic models by relaxing the rate-of-strain tensor. An open source finite element environment (i.e.~FreeFEM++\cite{freefem}) is used to discretise Eqs.~(\ref{eq:non-govern}) \& (\ref{eq:non-const}) and perform simulations with anisotropic adaptive mesh in $\Gamma / \bar{X}$. This entire computational approach has been extensively explained and validated in our previous studies; further details can be found in Refs.~[\hspace{-2pt} \citenum{chaparian2017yield,chaparian2019adaptive,iglesias2020computing,chaparian2021sliding,medina2023rheo}]. Sample simulations are shown in Fig.~\ref{fig:Torque_Bingham_axis}.

\subsection{3D computational method}

\begin{figure}
\includegraphics[width=\linewidth]{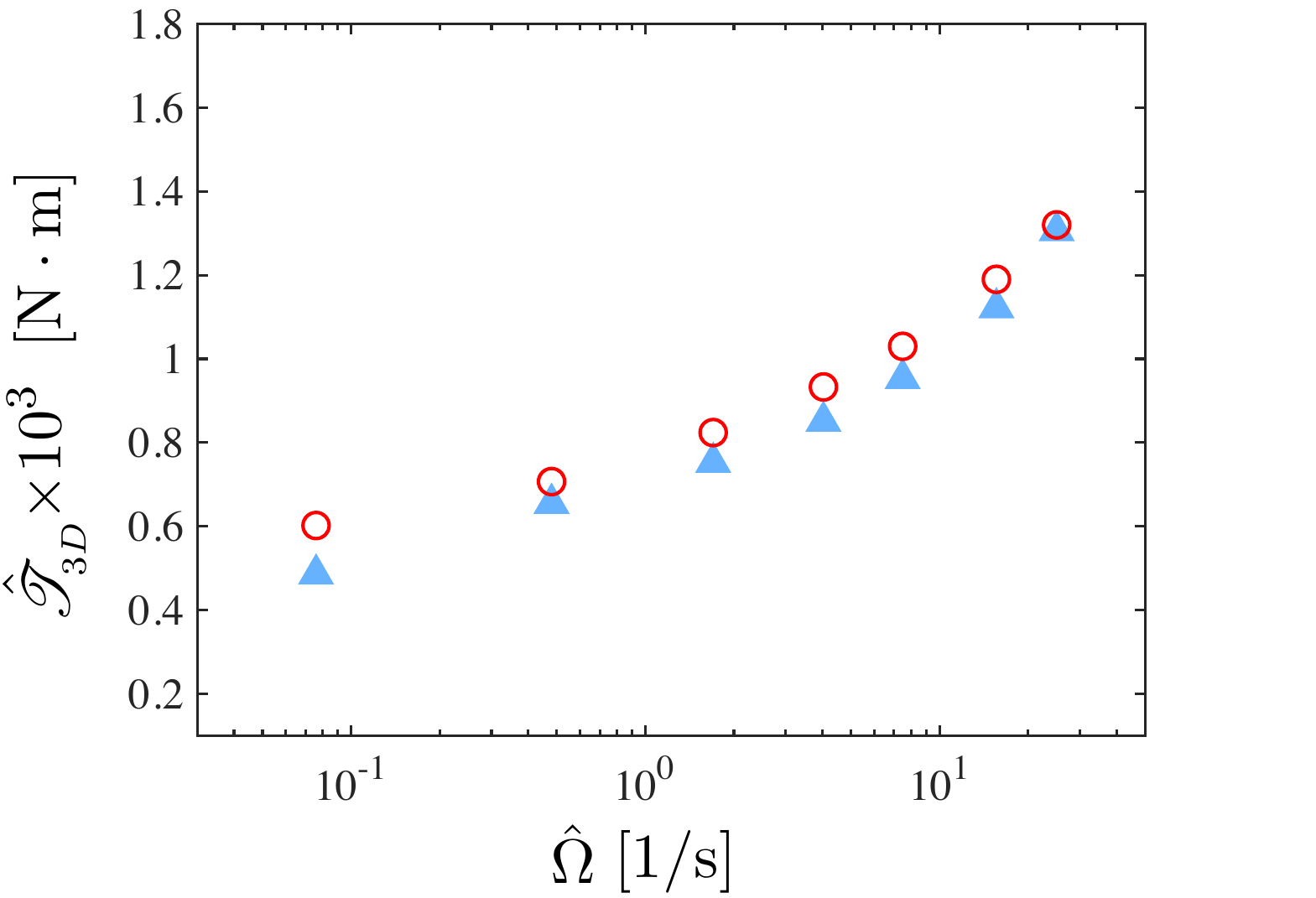}
\caption{Torque versus $\hat{\Omega}$ for a 6-arm vane fixture. The red circular symbols are the present 3D computational results and the blue triangles are the experimental data of \citet{medina2019rheo} for a Carbopol microgel; see Fig.~(4) in the reference.}
\label{fig:validation}
\end{figure}

Although the augmented Lagrangian method is generally more robust and accurate than the widely used regularisation approaches for simulating yield-stress fluid flows \cite{frigaard2005usage}, its high computational cost becomes prohibitive for 3D simulations, particularly for the complex fractal vane geometries considered in the present study. Consequently, in the present study, 3D computations are performed using a regularisation based finite volume solver implemented in OpenFOAM.

A steady-state finite volume solver is employed together with a body-fitted mesh predominantly composed of hexahedral cells, generated using \texttt{blockMesh} and \texttt{snappyHexMesh}. The former is used to generate the initial background mesh in $\Gamma$, while the latter incorporates the vane geometry ($\partial X$) and performs local mesh refinement. Additional refinement is introduced around the vane arms and within the yielded regions to accurately resolve the large velocity gradients. The vane rotation is modelled using the multiple reference frame (MRF) approach, which is appropriate for the steady-state flows considered here. For the narrow-gap configuration (case $\mathbb{A}$), approximately 40 cells are employed across the radial gap between the vane tips and the cup wall, whereas this number is increased to approximately 90 for the wide-gap configuration (case $\mathbb{B}$). In the axial direction, approximately 110 cells are used over the cup height, while the vane length is discretised using approximately 190 and 130 cells for cases $\mathbb{A}$ and $\mathbb{B}$, respectively.

The constitutive equation (\ref{eq:const}) is regularised using the well-known Papanastasiou formulation combined with a narrow bi-viscosity region to maintain the numerical stability in zones of extremely low shear rate \cite{jalaal2016long}. This regularisation provides a smooth transition between yielded and unyielded regions while closely approximating the original Herschel-Bulkley model. At the upper boundary of the 3D computational domain, corresponding to the free surface of the sample, a zero-traction boundary condition is imposed while surface deformation is neglected.

The accuracy of our 3D simulations is extensively assessed through standard CFD verification procedures, including grid-independence and regularisation parameters sensitivity, comparison of the computed velocity profiles with 1D analytical circular Couette solution (derived in Appendix\ref{sec:Couette}), and most importantly, validation against the experimental measurements of \citet{medina2019rheo}. Fig.~\ref{fig:validation} compares our computed torque (red circles) with the experimental measurements (blue triangles) reported in the reference for a 6-arm vane fixture, as a function of the vane rotation rate. The vanishingly small discrepancy between the measured and computed torques is primarily attributed to the boundary conditions. In the experiments of \citet{medina2019rheo}, the Carbopol microgel exhibited a finite degree of wall slip, whereas the present simulations assume no-slip conditions on both the vane and cup surfaces. As discussed in \S \ref{sec:intro}, our previous study demonstrated that slip at the vane surface has a negligible influence on the measured torque \cite{chaparian2022vane}. Slip at the cup wall, however, can affect the measured torque, particularly at low rotation rates. This trend is evident in Fig.~\ref{fig:validation}, where the agreement between the experimental and numerical results improves as the rotation rate $\hat{\Omega}$ increases.

\section{2D features \& flow curve calculations}\label{sec:2D}
In this section, we consider the 2D problem and analyse the computational results for various vane geometries. These simulations are subsequently used to perform computational rheometry and extract the flow curve of the sample.

\begin{figure}
\includegraphics[width=\linewidth]{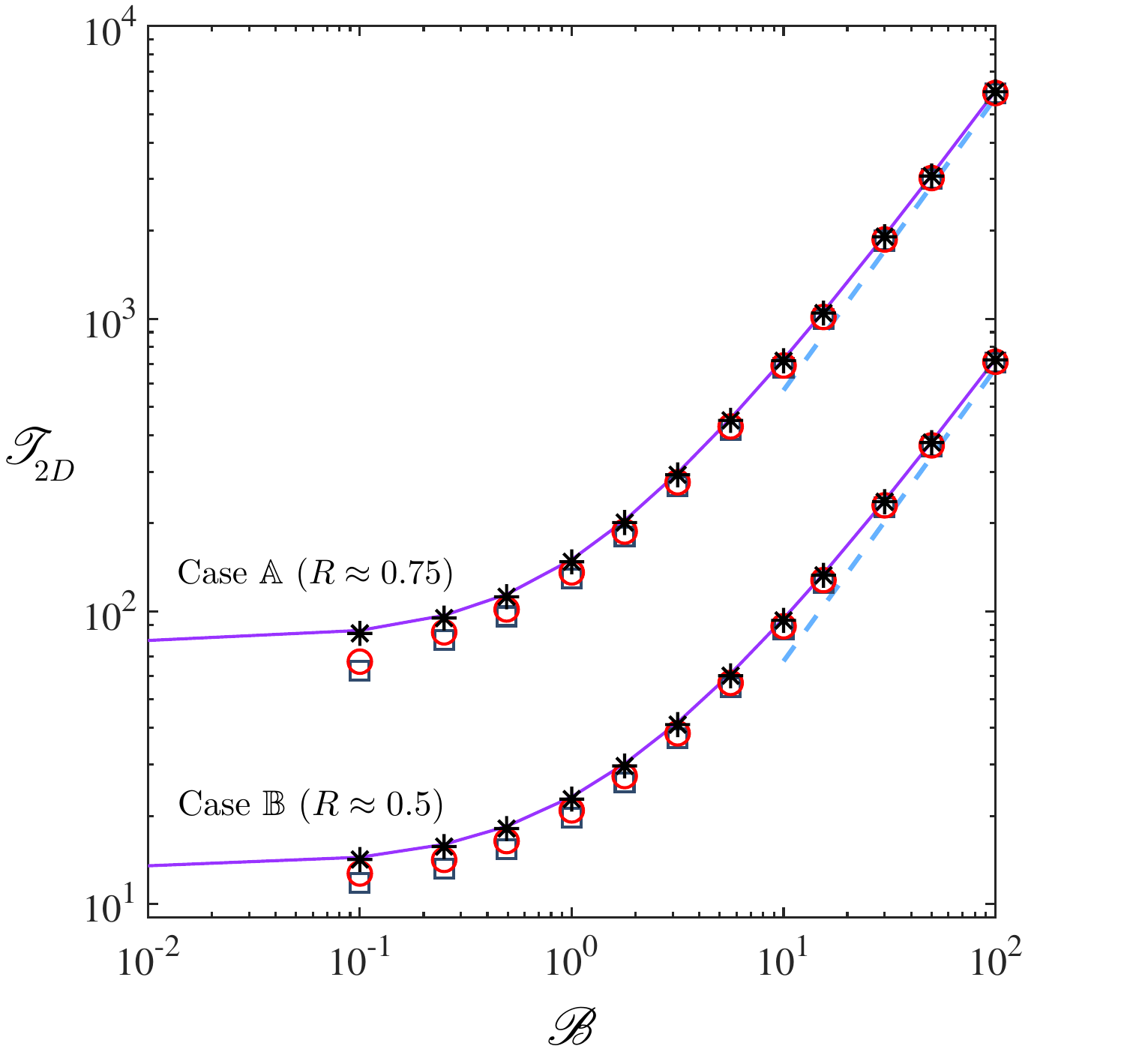}
\caption{Dimensionless torque versus the Bingham number for the 2D computations. The circular Couette solution (Appendix \ref{sec:Couette}), $\mathcal{T}_{2D}^c$, is shown with the continuous purple line and the yield/plastic asymptotic limit (expression (\ref{eq:asymptotic})) with the dashed light-blue line. The blue squares, red circles, and black asterisks are the computational data for the 4-arm, 6-arm, and 24-arm fractal vanes, respectively.}
\label{fig:2DTorque_Bingham}
\end{figure}

Fig.~\ref{fig:2DTorque_Bingham} shows the dimensionless torque $\mathcal{T}_{2D}$ as a function of the Bingham number for a sample Carbopol microgel characterised in Ref.~[\hspace{-4pt} \citenum{medina2023rheo}] with power-law index $n=0.46$. The continuous purple line represents the 1D circular Couette solution (Appendix \ref{sec:Couette}) which can be interpreted as a 2D equivalent of a bob \& cup rheometer. The blue squares, red circles, and black asterisks correspond to the computational results obtained using the 4-arm, 6-arm, and fractal vanes, respectively. Both configurations $\mathbb{A}$ and $\mathbb{B}$ are considered in this figure. Clear deviations from the purely azimuthal circular Couette flow are observed for the bespoke vanes, particularly in the narrow-gap configuration. In contrast, the fractal vane exhibits close agreement with the circular Couette solution in both configurations.

In the plastic limit ($\mathcal{B} \rightarrow \infty$), however, this discrepancy between various vane types disappears as the unyielded portion of the sample surrounding the vane effectively ``cloaks'' the actual vane geometry. At this limit of high Bingham numbers, plastic dissipation dominates the viscous one, so at the leading order, Eq.~(\ref{eq:energy_non}) reduces to,
\begin{equation}
\mathcal{B} \int_{\Gamma \setminus \bar{X}} \Vert \dot{\boldsymbol{\upgamma}} \Vert  ~\text{d}A \approx \mathcal{T}_{2D} \Big( \frac{1 - R}{R}\Big).
\end{equation}
In this limit, the sample is only yielded in a thin boundary layer adjacent to the vane tip; see Fig.~\ref{fig:Torque_Bingham_axis}(c). For circular Couette flow, this boundary layer forms an annular ring of thickness $\epsilon$ attached to the inner cylinder/bob. Hence, the shear rate in the yielded boundary layer scales as $\vert  \dot{\gamma}_{r\theta} \vert \approx 1/\epsilon$. Therefore,
\begin{equation}
\frac{\mathcal{B}}{\epsilon} 2\pi\frac{R}{1-R}\epsilon \approx \mathcal{T}_{2D}^c \Big( \frac{1 - R}{R} \Big),
\end{equation}
or indeed,
\begin{equation}\label{eq:asymptotic}
\mathcal{T}_{2D}^c \approx 2 \pi \Big( \frac{R}{1 - R} \Big)^2 \mathcal{B},
\end{equation}
which physically indicates that the shear stress on the bob surface asymptotically approaches the yield stress of the sample. Expression (\ref{eq:asymptotic}) is shown with a light blue dashed line in Fig.~\ref{fig:2DTorque_Bingham} which successfully captures the torque in $\mathcal{B} \rightarrow \infty$ limit. Away from this plastic limit, however, torque-to-stress and rotation-to-strain rate conversions are required for flow curve calculations.

For this aim, we take a closer look at the circular Couette case; see Appendix \ref{sec:Couette}. For this case, the only non-zero components of the stress and strain rate tensors are $\hat{\tau}_{r \theta}$ and $\hat{\dot{\gamma}}_{r \theta}$. For the sake of brevity, we drop $_{r\theta}$ in the following derivations. The shear stress and shear rate can be written as,
\begin{equation}\label{eq:stress_radius}
\frac{\text{d} \hat{r}}{\hat{r}} = - \frac{1}{2} \frac{\text{d} \hat{\tau}}{\hat{\tau}},
\end{equation}
and,
\begin{equation}\label{eq:gammadot}
\hat{\dot{\gamma}} = \hat{r} \frac{\text{d}}{\text{d}\hat{r}} \left( \frac{\hat{u}_{\theta}}{\hat{r}} \right) = \hat{r} \frac{\text{d} \hat{\omega}}{\text{d}\hat{r}},
\end{equation} 
where $\hat{\omega}=\hat{u}_{\theta}/\hat{r}$ is the fluid's rotation. The constitutive equation (\ref{eq:const}) can be reformed as,
\begin{equation}
\hat{\dot{\boldsymbol{\upgamma}}} = f(\hat{\boldsymbol{\uptau}}) = \left( \frac{ \Vert \hat{\boldsymbol{\uptau}} \Vert - \hat{\tau}_y }{ \hat{K} \Vert \hat{\boldsymbol{\uptau}} \Vert^n } \right)_+^{1/n} ~\hat{\boldsymbol{\uptau}} .
\end{equation}
Substituting expression (\ref{eq:stress_radius}) into (\ref{eq:gammadot}),
\begin{equation}
\hat{\dot{\gamma}} = f(\hat{\tau}) = \hat{r} \frac{\text{d} \hat{\omega}}{\text{d}\hat{r}} = -2 \hat{\tau}\frac{\text{d} \hat{\omega}}{\text{d} \hat{\tau}}.
\end{equation}
By integrating the fluid's rotation from the inner cylinder to the stationary cup,
\begin{equation}
- \int_{\hat{\Omega}}^0 \text{d}\hat{\omega} = \hat{\Omega} = \frac{1}{2} \int_{\hat{\tau}_1}^{\hat{\tau}_2} \frac{f(\hat{\tau})}{\hat{\tau}} \text{d} \hat{\tau},
\end{equation}
we can find the relation between the applied rotation rate and the shear stress in the gap. Derivative of the above equation with respect to $\hat{\tau}_1$ yields,
\begin{equation}
\frac{\text{d} \hat{\Omega}}{\text{d} \hat{\tau}_1} = \frac{1}{2} \left[ \frac{f(\hat{\tau}_2)}{\hat{\tau}_2} \frac{\text{d} \hat{\tau}_2}{\text{d} \hat{\tau}_1} - \frac{f(\hat{\tau}_1)}{\hat{\tau}_1} \right].
\end{equation}
Knowing that $\hat{\tau}_2 = \left( \displaystyle \frac{\hat{R}_1}{\hat{R}_2} \right)^2 \hat{\tau}_1 = R^2 \hat{\tau}_1 $ (see Eq.~(\ref{eq:stress_decay}) in Appendix \ref{sec:Couette}),
\begin{equation}
\frac{\text{d} \hat{\Omega}}{\text{d} \hat{\tau}_1} = \frac{1}{2 \hat{\tau}_1} \left[ f(\hat{\tau}_2) - f(\hat{\tau}_1) \right],
\end{equation}
which can be reformed as,
\begin{equation}
\frac{2 ~\text{d}\hat{\Omega}}{ \displaystyle \frac{\text{d} \hat{\tau}_1}{\hat{\tau}_1} } = \frac{2 ~\text{d}\hat{\Omega}}{ \text{d} (\log \hat{\tau}_1) } = f(\hat{\tau}_2) - f(\hat{\tau}_1).
\end{equation}
Knowing that $\text{d} \hat{\Omega} = \hat{\Omega} ~\text{d} \left( \log \hat{\Omega} \right)$, the above equation can be rewritten as,
\begin{equation}\label{eq:gamma_full}
\frac{2 \hat{\Omega} }{ \displaystyle \frac{\text{d} \left( \log \hat{\tau}_1 \right)}{ \text{d} \left( \log \hat{\Omega} \right) } } = \frac{2 \hat{\Omega} }{ \displaystyle \frac{\text{d} \left( \log \hat{\mathcal{T}} \right)}{ \text{d} \left( \log \hat{\Omega} \right) } } = f(\hat{\tau}_2) - f(\hat{\tau}_1).
\end{equation}
Since our focus in the present work is on yield-stress fluids, we assume $f(\hat{\tau}_2) \approx 0$, or more generally $\vert f(\hat{\tau}_2) \vert \ll \vert f(\hat{\tau}_1) \vert$. Physically, this implies that the fluid adjacent to the cup wall is either unyielded or undergoes significantly less deformation than the fluid near the bob surface. This assumption is throuughly valid in the partially yielded regime or at low rotation rates, $\hat{\Omega}$. In Appendix~\ref{sec:conversion}, we assess the validity of this approximation over the broad range of effective parameters considered in the present study, using one-dimensional circular Couette flow calculations. Under this assumption, expression~(\ref{eq:gamma_full}) simplifies to,
\begin{equation}\label{eq:wide_gap}
\hat{\dot{\gamma}}_1 = \frac{-2 \hat{\Omega} }{ \displaystyle \frac{\text{d} \left( \log \hat{\mathcal{T}} \right)}{ \text{d} \left( \log \hat{\Omega} \right) } },
\end{equation}
which has been widely adopted in the literature as a practical relation for converting the vane rotation rate into the corresponding shear strain rate.

Regarding the torque-to-stress conversion, \citet{chaparian2022vane} reviewed previously proposed expressions and their origins \cite{sherwood1991use,atkinson1992torque}. In that reference, the authors used,
\begin{equation}\label{eq:2D_stress_conversion}
\hat{\tau} = \frac{\hat{\mathcal{T}}_{2D}}{2\pi \hat{R}_1^2 \left( 1-\displaystyle\frac{C}{N} \right) },
\end{equation}
which accounts for the effects of non-uniform shear and the combined extensional and shear deformations exerted by the vanes on the sample. In this expression, $C=1.113$ and $N$ is the number of vane arms \cite{atkinson1992torque}. In the case of a bob geometry (i.e.~$N \to \infty$), expression (\ref{eq:2D_stress_conversion}) reduces to $\hat{\mathcal{T}}_{2D}=2 \pi \hat{R}_1^2 \hat{\tau}$ (i.e.~expression (\ref{eq:Couette_torque_stress})) which corresponds to the classical circular Couette relation.

\begin{figure*}
\centerline{\includegraphics[width=\linewidth]{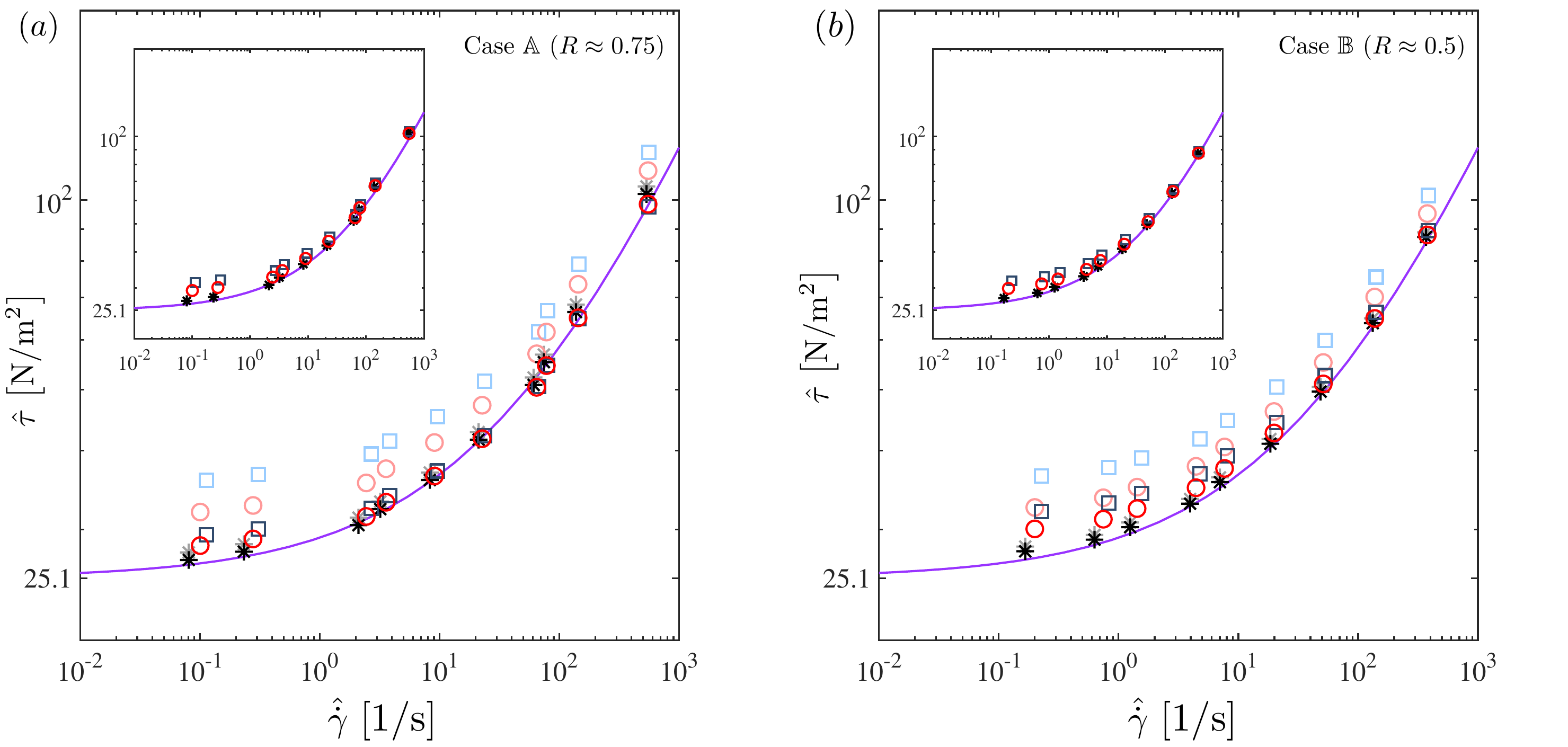}}
\caption{Calculated flow curves (shear stress versus shear rate) from the 2D computational results: (a) case $\mathbb{A}$ and (b) case $\mathbb{B}$. The actual flow curve ($\hat{\tau}_y=25.1 ~\text{Pa}, \hat{K}=4 ~\text{Pa}\cdot \text{s}^\text{n} ~\&~ n=0.46$) is shown by the continuous purple line. The blue squares, red circles, and black asterisks correspond to the 4-arm, 6-arm, and 24-arm fractal vanes data, respectively. The darker symbols are calculated using our fitted parameters: (a) $C_{\text{\it opt}}^{\mathbb{A}}$ and (b) $C_{\text{\it opt}}^{\mathbb{B}}$. The lighter symbols represent the converted values using the original parameter. Insets: each inset shows the data of the corresponding panel converted with the averaged parameter. The parameters are reported in Table \ref{tab:conversion_parameters}.}
\label{fig:2D_flowcurve}
\end{figure*}

\setlength{\tabcolsep}{10pt} % Default value: 6pt
\renewcommand{\arraystretch}{1.5} % Default value: 1
\begin{table*}[t]
%\begin{ruledtabular}
\begin{tabular}{ c c c c c }
\hline
\hline
\colrule
Original\cite{chaparian2022vane} & $C=1.113$ & $D=2.75$ & $E=3$ & --- \\
\hline
Case $\mathbb{A}$  & $C_{\text{\it opt}}^{\mathbb{A}} \approx 0.480$  & $D_{\text{\it opt}}^{\mathbb{A}} \approx 3.722$  & $E_{\text{\it opt}}^{\mathbb{A}} \approx 4.204$ & $R-squ. \approx 0.98629$ \\
\hline 
Case $\mathbb{B}$  & $C_{\text{\it opt}}^{\mathbb{B}} \approx 0.715$  & $D_{\text{\it opt}}^{\mathbb{B}} \approx 3.197$  & $E_{\text{\it opt}}^{\mathbb{B}} \sim O(10^{-2})$ & $R-squ. \approx 0.98205$ \\
\hline
Mean values & $C_{\text{\it opt}}^{mean} \approx 0.686 $  & $D_{\text{\it opt}}^{mean} \approx 3.321$  & $E_{\text{\it opt}}^{mean} \approx 1.519$  & $R-squ. \approx 0.98245$ \\
\hline
\hline
\end{tabular}
%\end{ruledtabular}
\caption{\label{tab:conversion_parameters}%
Parameters for torque-to-stress conversion.}
\end{table*}

We now have a complete set of tools (i.e.~expressions (\ref{eq:wide_gap}) and (\ref{eq:2D_stress_conversion})) to convert the data presented in Fig.~\ref{fig:2DTorque_Bingham} into the corresponding flow curves. For this aim, we consider again the Carbopol microgel sample from Ref.~[\hspace{-4pt} \citenum{medina2023rheo}] with rheological properties $\hat{\tau}_y=25.1 ~\text{Pa}, \hat{K}=4 ~\text{Pa} \cdot \text{s}^{\text{n}} ~\&~ n=0.46$. Fig.~\ref{fig:2D_flowcurve} compares the calculated flow curves (the symbols) with the actual one (purple line). The same as Fig.~\ref{fig:2DTorque_Bingham}, the blue squares, red circles, and black asterisks correspond to the 4-arm, 6-arm, and fractal vanes data, respectively. The lighter symbols represent the converted values using $C=1.113$ in expression (\ref{eq:2D_stress_conversion}). A substantial discrepancy is evident for the conventional bespoke vanes. We therefore determine an optimal correction factor, $C_{\text{\it opt}}$, by minimising the deviation between the calculated and prescribed flow curves. The resulting predictions, shown by the darker symbols, exhibit excellent agreement with the constitutive equation. However, the choice of $C$ has only a minor influence on the flow curve obtained using the 24-arm fractal vane. This behaviour is expected because the fractal vane generates an almost purely azimuthal flow field (see Fig.~\ref{fig:Torque_Bingham_axis}(a-c)), making the correction term $C/N$ largely unnecessary for the torque-to-stress conversion. The optimised values are reported in Table~\ref{tab:conversion_parameters} for both the narrow- and wide-gap configurations. As expected, the optimised parameter depends on the radius ratio, increasing as $R$ decreases. Nevertheless, a representative mean value can be obtained by simultaneously considering cases $\mathbb{A}$ and $\mathbb{B}$. The insets in Fig.~\ref{fig:2D_flowcurve} show the flow curves reconstructed using this mean parameter $C_{\text{\it opt}}^{mean}$, demonstrating good agreement with the prescribed flow curve for both configurations.

\section{3D effects}\label{sec:3D}

\begin{figure*}
\centerline{\includegraphics[width=\linewidth]{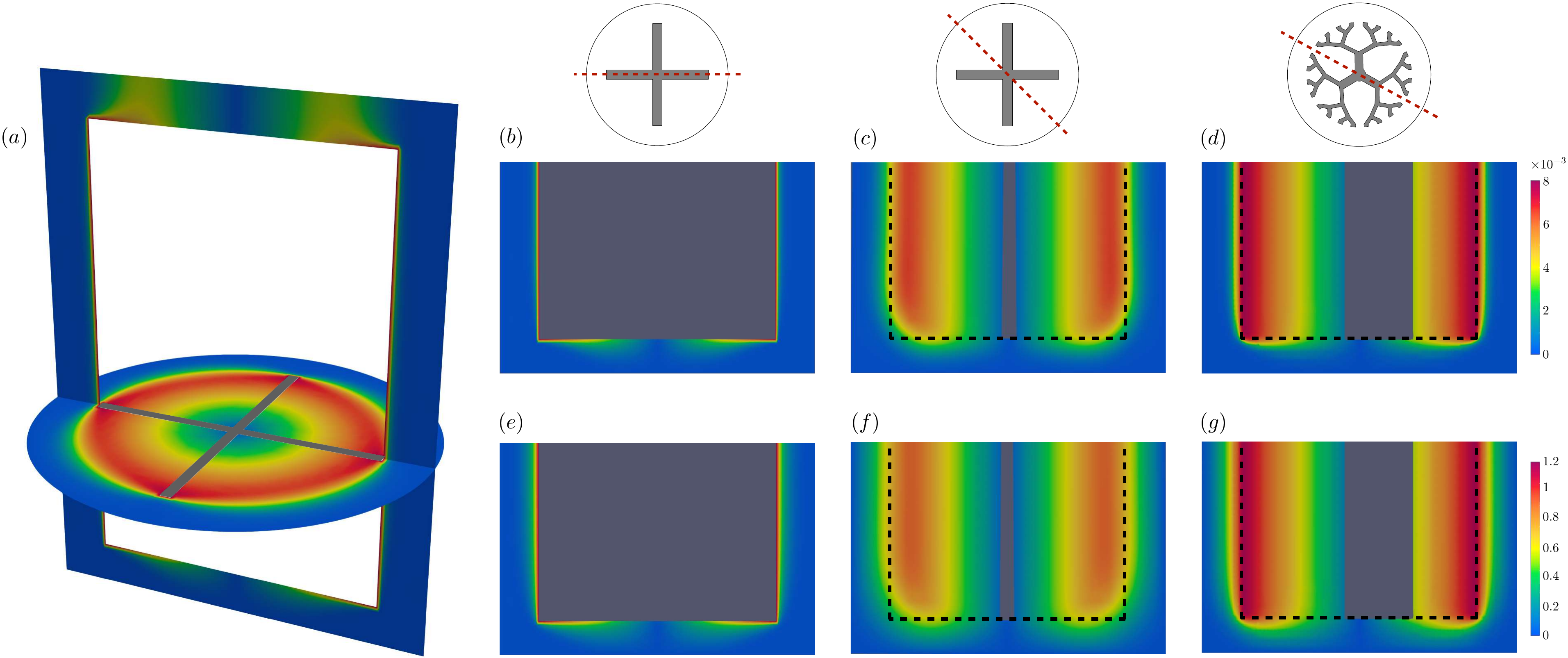}}
\caption{The velocity magnitude $\vert \hat{\boldsymbol{u}} \vert$ contours over various azimuthal planes (indicated by the red dashed lines in the top schematics). Panels (a,b,c,e,f) correspond to the 4-arm and panels (d,g) to the 24-arm fractal vanes of type $\mathbb{A}$. In panels (a-d), $\hat{\Omega}=0.57 ~\text{s}^{-1}$, while in panels (e-g), $\hat{\Omega}=83.87~\text{s}^{-1}$. Only the bottom half of the vane-in-cup fixture is shown in panels (b-g). The dashed black lines in panels (c,d,f,g) show the revolution boundary of the vane. In all cases, the sample has identical rheological properties: $\hat{\tau}_y=25.1 ~\text{Pa}, \hat{K}=4 ~\text{Pa}\cdot \text{s}^\text{n} ~\&~ n=0.46$.}
\label{fig:vel_contour}
\end{figure*}

\begin{figure}
\centerline{\includegraphics[width=\linewidth]{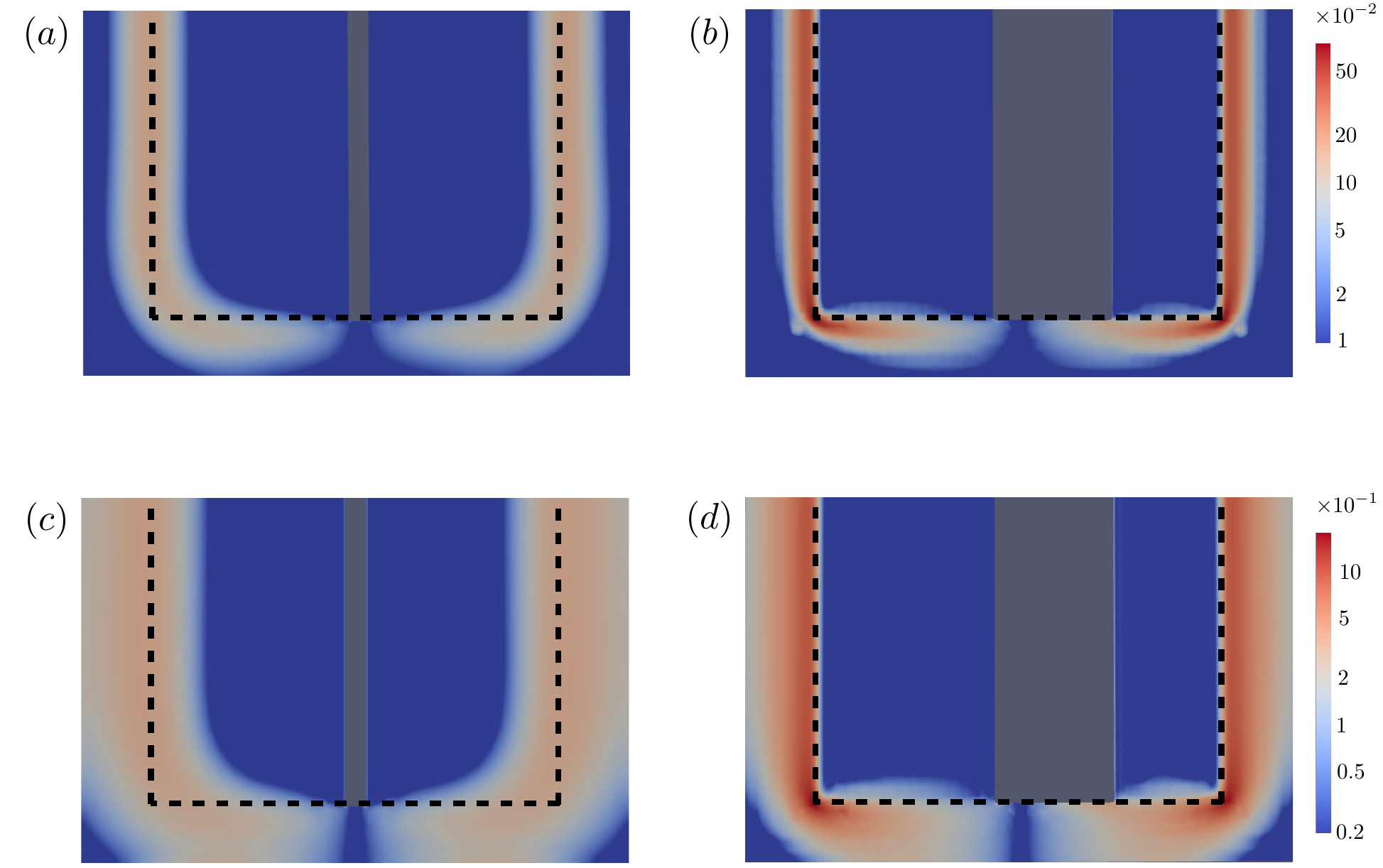}}
\caption{The contours represent $\Vert \hat{\boldsymbol{\dot{\upgamma}}} \Vert$ scaled with the nominal shear rates. This figure corresponds to the simulations presented in Fig.~\ref{fig:vel_contour}: panels (a,c) to Fig.~\ref{fig:vel_contour}(c,f) and panels (b,d) to Fig.~\ref{fig:vel_contour}(d,g). The nominal shear rates are: (a) 17.581 $1/\text{s}$, (b) 622.552 $1/\text{s}$, (c) 14.685 $1/\text{s}$, and (d) 545.034 $1/\text{s}$. Note that the scale of the colourbar is logarithmic.}
\label{fig:gammadot_contour}
\end{figure}

Having addressed the 2D problem, we now investigate the 3D effects in this section. 

Sample 3D simulations are illustrated in Fig.~\ref{fig:vel_contour} for the 4-arm bespoke and 24-arm fractal vanes of type $\mathbb{A}$. The contours represent the velocity magnitude $\vert \hat{\boldsymbol{u}} \vert$ on selected planes within the computational domain. In panel (a) and all the top row panels, $\hat{\Omega} = 0.57 ~s^{-1}$, while the bottom row panels correspond to $\hat{\Omega}=83.87~s^{-1}$. Consistent with the 2D results, the fractal vane produces flow kinematics that are much closer to the ideal bob \& cup configuration, characterised by nearly pure azimuthal motion. This is particularly evident by comparing panels (d,g) with panels (c,f). Although all four panels show the velocity field on the azimuthal plane passing through the middle of the vane arms, the maximum velocity in the fractal vane configuration is concentrated around the circumferential surface indicated by the black dashed line. This shows that the fluid enclosed by the fractal arms is effectively ``cloaked'' and rotates almost as a rigid body. In contrast, the conventional 4-arm vane generates a highly non-uniform flow, with significant variations in both the radial and axial directions: there is a large difference between the velocity field on the mid $z$-plane and that near the bottom of the vane. Another notable feature is that the yielded region beneath the vane extends further towards the bottom of the cup for the fractal geometry, increasing the volume of sheared material and consequently the measured torque. These differences are more clearly illustrated in Fig.~\ref{fig:gammadot_contour}, which presents contours of the scaled strain rate magnitude $\Vert \hat{\boldsymbol{\dot{\upgamma}}} \Vert$. For the fractal vane (panels (b) and (d)), nearly the entire fluid region enclosed by the dashed black lines remains unyielded, even at high rotation rates. In contrast, for the 4-arm vane, the yielded region extends into the area enclosed by the dashed black lines, further confirming the non-circular nature of the yield surface observed in Fig.~\ref{fig:Torque_Bingham_axis}d.

\begin{figure*}
\includegraphics[width=\linewidth]{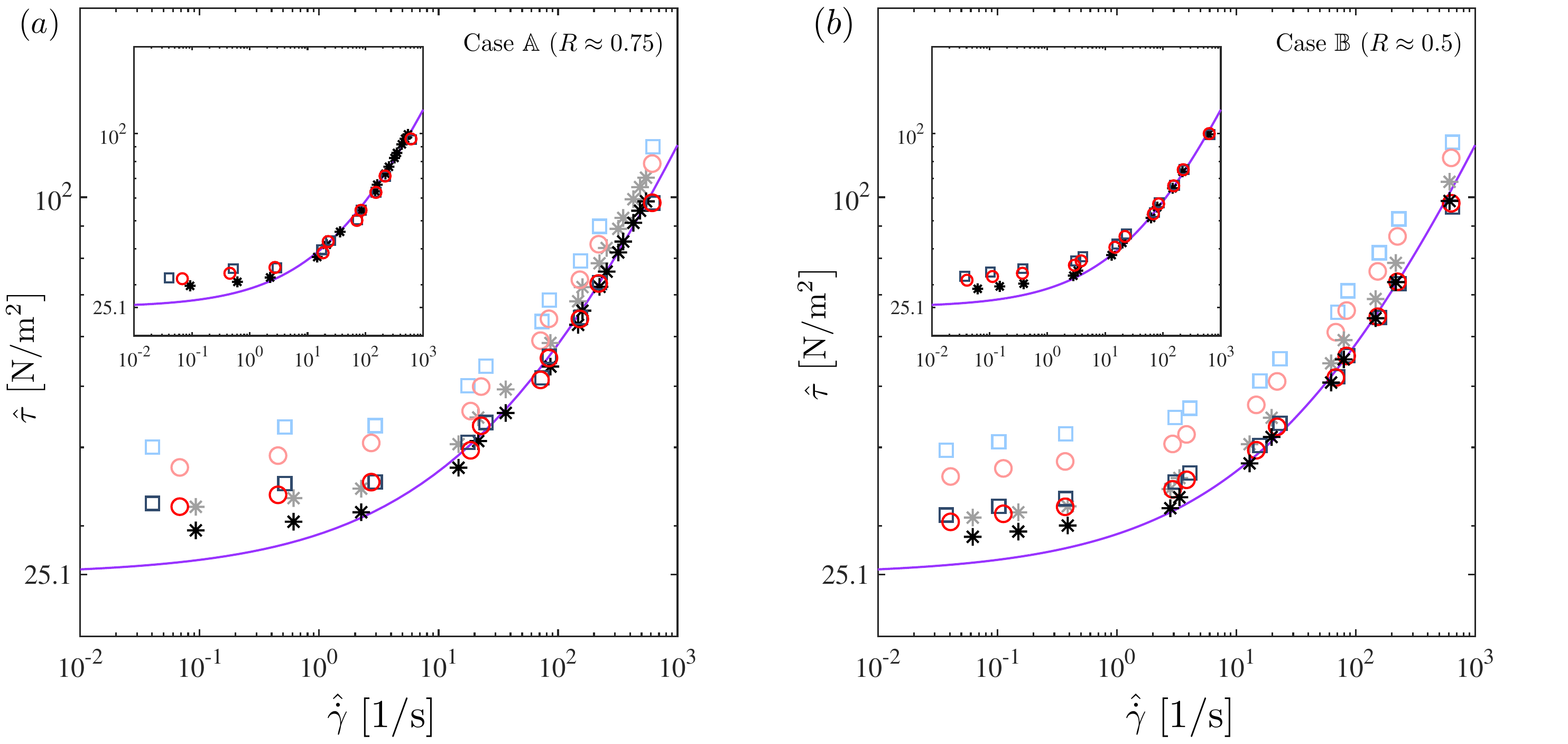}
\caption{Calculated flow curves (shear stress versus shear rate) from the 3D computational results: (a) case $\mathbb{A}$ and (b) case $\mathbb{B}$. The actual flow curve ($\hat{\tau}_y=25.1 ~\text{Pa}, \hat{K}=4 ~\text{Pa}\cdot \text{s}^\text{n} ~\&~ n=0.46$) is shown by the continuous purple line. The blue squares, red circles, and black asterisks correspond to the 4-arm, 6-arm, and 24-arm fractal vanes data, respectively. The darker symbols are calculated using our optimised parameters: 
$C_{\text{\it opt}}^{\mathbb{A}}$, $D_{\text{\it opt}}^{\mathbb{A}}$ \& $E_{\text{\it opt}}^{\mathbb{A}}$ and (b) $C_{\text{\it opt}}^{\mathbb{B}}$, $D_{\text{\it opt}}^{\mathbb{B}}$ \& $E_{\text{\it opt}}^{\mathbb{B}}$. The lighter symbols represent the converted values using the original parameter. Insets: each inset shows the data of the corresponding panel converted with the averaged parameters. The parameters are reported in Table \ref{tab:conversion_parameters}.}
\label{fig:3D_flowcurve}
\end{figure*}

Further insights into the 3D features can be obtained by examining the flow curves for each vane geometry. To convert the experimentally measured torque in a vane-in-cup rheometer into the corresponding shear stress, \citet{medina2019rheo} employed the relation,
\begin{equation}
\hat{\tau} = \frac{\mathcal{\hat{T}}_{3D}}{2\pi \hat{R}_{\text{\it eff}}^2 \hat{L} \left( 1+\displaystyle\frac{2\hat{R}_{\text{\it eff}}}{3\hat{L}} \right) }
\end{equation}
in which $\hat{R}_{\text{\it eff}}$ denotes the ``effective radius of the vane''. Physically, $\hat{R}_{\text{\it eff}}$ corresponds to the radial location at which the fluid reaches its maximum linear velocity. However, it has since been shown that this is not an ideal choice, as $\hat{R}_{\text{\it eff}}$ itself depends on the rheological properties of the sample. To overcome this limitation, \citet{chaparian2022vane} used,
\begin{equation}\label{eq:3D_stress_conversion}
\hat{\tau} = \frac{\hat{\mathcal{T}}_{3D}}{2\pi \hat{R}_1^2 \hat{L} \left[ \left( 1-\displaystyle\frac{C}{N} \right) + \displaystyle \frac{\hat{R}_1}{4\hat{L}} \left( D - \displaystyle \frac{E}{\sqrt{N}} \right) \right] }
\end{equation}
which is indeed the 3D extension of expression (\ref{eq:2D_stress_conversion}). As mentioned in \S \ref{sec:2D}, the original value of $C=1.113$; moreover, $D=2.75$ and $E=3$. The first term in the denominator is identical to that in expression (\ref{eq:2D_stress_conversion}) and accounts for the mixed state of deformation (shear and extension) generated by the vane arms. The second term, however, provides a correction for the 3D end effects. In the previous section, we showed that the agreement between the calculated flow curves and the actual one can be substantially improved by replacing the original $C$ with the optimised values. Keeping this value fixed at the optimised value, we now investigate whether further improvements can be achieved by optimising the remaining parameters $D$ and $E$.

Figure~\ref{fig:3D_flowcurve} compares the flow curves calculated from the 3D simulations (symbols) with the prescribed Herschel-Bulkley fluid (solid purple line). The symbols follow the same convention as in Fig.~\ref{fig:2D_flowcurve}: the lighter symbols correspond to the original values of $C$, $D$, and $E$, whereas the darker symbols are obtained using the optimised parameters reported in Table \ref{tab:conversion_parameters}. The same trend observed in the 2D results is evident here as well. The 24-arm fractal vane already provides a great agreement with the prescribed flow curve using the original parameters, with only a marginal improvement after optimisation. In contrast, the calculated flow curves for the conventional bespoke vanes exhibit substantially larger discrepancies, which are significantly reduced by employing the optimised parameters. The insets in Fig.~\ref{fig:3D_flowcurve} show the computational data converted using the mean parameters. As in Fig.~\ref{fig:2D_flowcurve} insents, good agreement with the prescribed constitutive flow curve is obtained for both configurations, demonstrating the effectiveness of the mean conversion parameters.

\begin{figure}
\centerline{\includegraphics[width=0.95\linewidth]{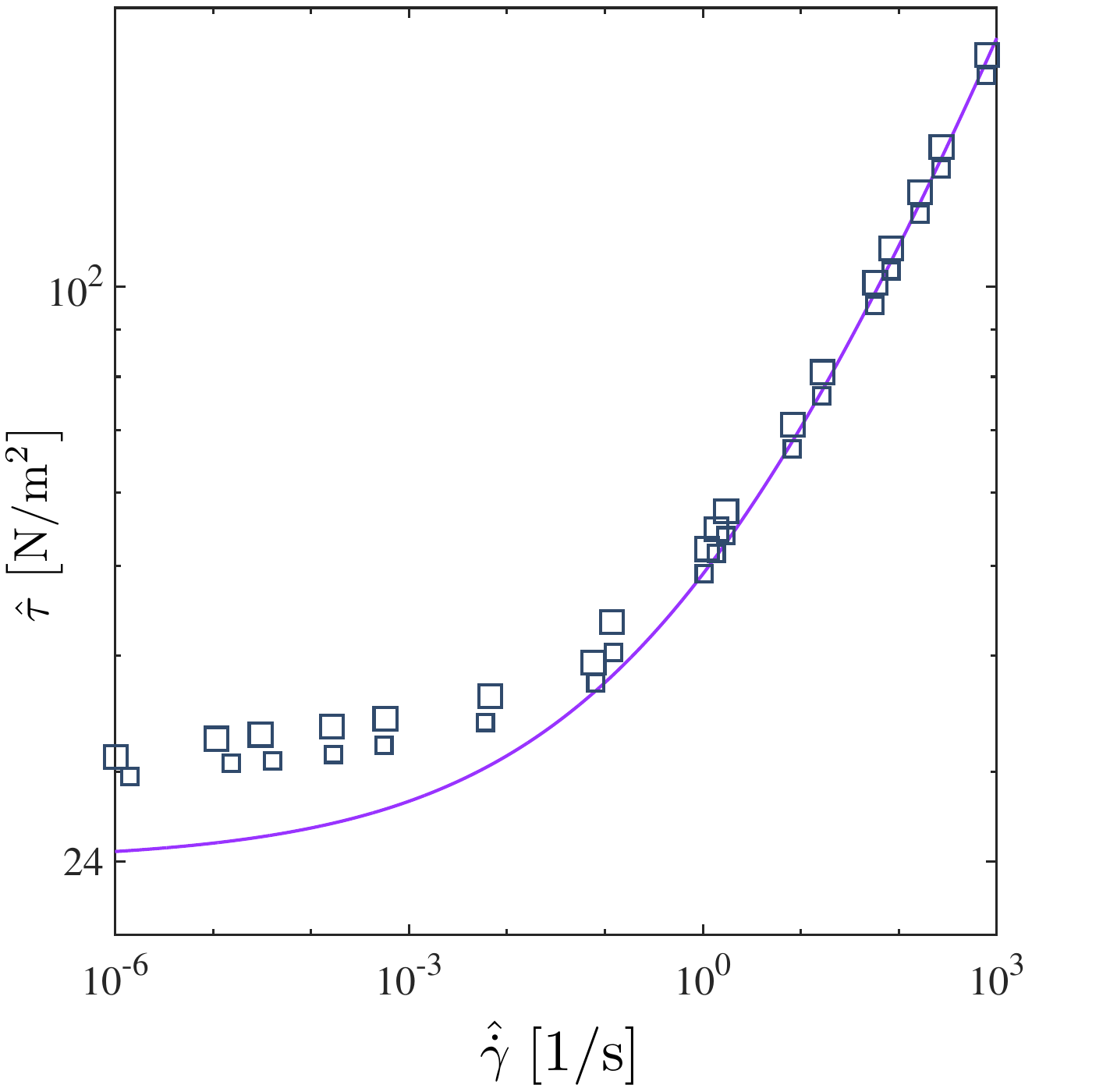}}
\caption{Calculated flow curves (shear stress versus shear rate) from the 3D computational results. The actual flow curve ($\hat{\tau}_y=24 ~\text{Pa}, \hat{K}=25 ~\text{Pa}\cdot \text{s}^\text{n} ~\&~ n=0.27$) is shown by the continuous purple line. The blue squares correspond to the 4-arm vanes data: smaller symbols to the narrow gap (case $\mathbb{A}$) and larger symbols to the wide gap (case $\mathbb{B}$), respectively. Both symbols are calculated with averaged parameters in Table \ref{tab:conversion_parameters}.}
\label{fig:shavingfoam}
\end{figure}

\section{Summary \& conclusion}\label{sec:conclusion}

In this study, we developed a computational framework capable of simulating 3D kinematics of flows within vane-in-cup rheometers. More specifically, we studied the behaviour of a Herschel-Bulkley viscoplastic fluid tested with various vane geometries, including 4- and 6-arm bespoke vanes and a 24-arm fractal vane both in narrow- and wide-gap configurations.

As a baseline, in \S \ref{sec:2D}, we first examined the features of the flow in a 2D set-up. Anisotropic adaptive finite element method based on augmented Lagrangian scheme was employed to simulate the flow within the cross section of the vane-in-cup geometry. Recently, \citet{chaparian2022vane}, using exactly the same method, investigated the kinematics, yielding behaviour, and relative contributions of shear and extensional deformation rates for various $N$-arm bespoke and fractal vanes. The present study rather focuses primarily on the direct implications of vane geometry for rheological measurements, specifically on how the raw rheometry data (vane rotation rate and applied torque) should be converted into shear stress and strain rate to minimise errors in the extracted rheological properties.

Subsequently, in \S \ref{sec:3D}, we extended the analysis to fully 3D simulations and proposed an optimised stress conversion closure. To the best of our knowledge, this work provides the first detailed characterisation of the 3D kinematics of a viscoplastic sample within a vane-in-cup rheometer. The results demonstrate how the vane geometry controls the yielded region and the resulting flow structure. In particular, the 4-arm vane generates a strongly non-uniform flow field with substantial deviations from circular Couette kinematics in both the radial and axial directions. In contrast, the 24-arm fractal vane produces a flow field that is effectively ``cloaked'' by the special structure of the arms, resulting in a much more uniform azimuthal shear. Consequently, the fractal vane exhibits significantly closer similarities to the classical bob \& cup geometry and provides a stronger physical basis for the use of the circular Couette analogy in rheological interpretation.

Although the preceding analysis focused on a Carbopol microgel with rheological properties characterised in Ref.~[\hspace{-4pt} \citenum{medina2023rheo}], the improved conversion relations are not restricted to this particular material. To demonstrate their broader applicability, we have performed additional simulations for a highly shear-thinning shaving foam characterised by a different set of rheological parameters reported in Ref.~[\hspace{-4pt} \citenum{ovarlez2010investigation}]. Fig.~\ref{fig:shavingfoam} compares the flow curves reconstructed using the mean optimised conversion parameters with the prescribed flow curve. The similar agreement between the computed flow curves and prescribed one as Fig.~\ref{fig:3D_flowcurve} demonstrates that the improved conversion relations remain effective for a material with substantially different rheological properties. These results further support the robustness of the proposed methodology and suggest that its applicability extends beyond the specific rheological parameters.

The present study provides new insights into computational rheometry and the quantification of rheological properties of yield-stress fluids using vane-in-cup fixtures. Nevertheless, several limitations remain, which provide opportunities for future investigations. First, complementary experimental measurements would further assess the applicability of the proposed conversion relations. Second, using the augmented Lagrangian approach for 3D simulations would provide a more accurate description in the yielding limit, although such computations remain prohibitively expensive with current computational resources. This is particularly important in low strain rate regimes. Furthermore, incorporating more advanced constitutive models \cite{saramito2009new,dimitriou2014comprehensive,dimitriou2018canonical,kamani2021unification} capable of describing thixo-elasto-visco-plastic behaviour would enable investigations of more complex phenomena, such as potential shear-banding observed in vane-in-cup measurements \cite{medina2023rheo,vadillo2025vane}. Other promising directions include studying the influence of loading history and vane immersion process, as well as systematically examining the effects of geometric parameters, such as gap width, bottom clearance, and vane/cup dimensions, on vane-in-cup rheometry.

\section*{Acknowledgments}

J.~M.~and E.~C.~acknowledge the support of University of Strathclyde via International Strategic Partner (ISP) Joint PhD Clusters. We thank Esteban F.~Medina-Bañuelos and José Pérez-González for sharing their data with us. Also, the high performance computational resources provided by ARCHIE-WeSt (Research Computing for the West of Scotland) for running 3D simulations are highly acknowledged.

\vspace{10pt}

{\bf Data availability.} The data that support the findings of this study are available from the corresponding author upon reasonable request.

%%%%%%%%%%%%%%%%%%%%%%%%%%%%%%%%%%

\appendix

\section{Couette analogy}\label{sec:Couette}

Here, we solve a 1D azimuthal flow between two concentric cylinders known as circular Couette flow where the inner cylinder rotates with $\hat{\Omega}$ and the outer cylinder is stationary. Hence, the Cauchy equation reduces to,
\begin{equation}
\frac{\text{d}(\hat{r}^2 \hat{\tau}_{r \theta})}{\text{d} \hat{r}}=0,
\end{equation}
which means that the shear stress in the gap is,
\begin{equation}\label{eq:stress_decay}
\hat{\tau}_{r\theta} = \frac{\hat{A}}{\hat{r}^2},
\end{equation}
where $\hat{A}$ is the constant of integration. The fluid behaviour obeys Eq.~(\ref{eq:const}), in which,
\begin{equation}
\hat{\dot{\gamma}}_{r\theta} = \hat{r} \frac{\text{d}}{\text{d}\hat{r}} \left( \frac{\hat{u}_{\theta}}{\hat{r}} \right),
\end{equation}
where $\hat{u}_{\theta}$ is the azimuthal velocity. Hence,
\begin{equation}\label{eq:Couette_gammadot}
  \left\{
    \begin{array}{ll}
      \displaystyle\frac{\hat{A}}{\hat{r}^2} = \hat{K}  ~\left\lvert \hat{\dot{\gamma}}_{r\theta}  \right\rvert ^{n-1} \hat{\dot{\gamma}}_{r\theta} + \hat{\tau}_y ~\text{sgn}(\hat{\dot{\gamma}}_{r\theta}) & \mbox{iff}\quad r < R_o, \\[10pt]
      \hat{\dot{\gamma}}_{r\theta} = 0 & \mbox{iff}\quad r \geqslant R_o.
  \end{array} \right.
\end{equation}
where $\hat{R}_o=\min \left( \sqrt{\vert \hat{A} \vert/\hat{\tau}_y} , \hat{R}_2 \right)$ is either the locus of the yield surface in the case of a partially yielded gap or the radius of the outer cylinder (i.e.~cup) in the case of a fully yielded gap, depending on the applied stress on the inner cylinder (indeed $\vert \hat{A} \vert/\hat{\tau}_y$ ratio). Indeed, the boundary conditions are $\hat{u}_{\theta}(\hat{R}_1)=\hat{R}_1 \hat{\Omega}$ and $\hat{u}_{\theta} (\hat{R}_o)=0$.
To find $\hat{u}_{\theta}$, we numerically integrate Eq.~(\ref{eq:Couette_gammadot}). The torque (per unit length) on the inner cylinder is,
\begin{equation}\label{eq:Couette_torque_stress}
\hat{\mathcal{T}}_{2D}^c = 2 \pi \hat{R}_1^2 \hat{\tau}_1 = 2 \pi \hat{A} .
\end{equation}

\section{Narrow \& wide gap rotation-to-strain rate conversion}\label{sec:conversion}

\begin{figure*}
\centerline{\includegraphics[width=0.9\linewidth]{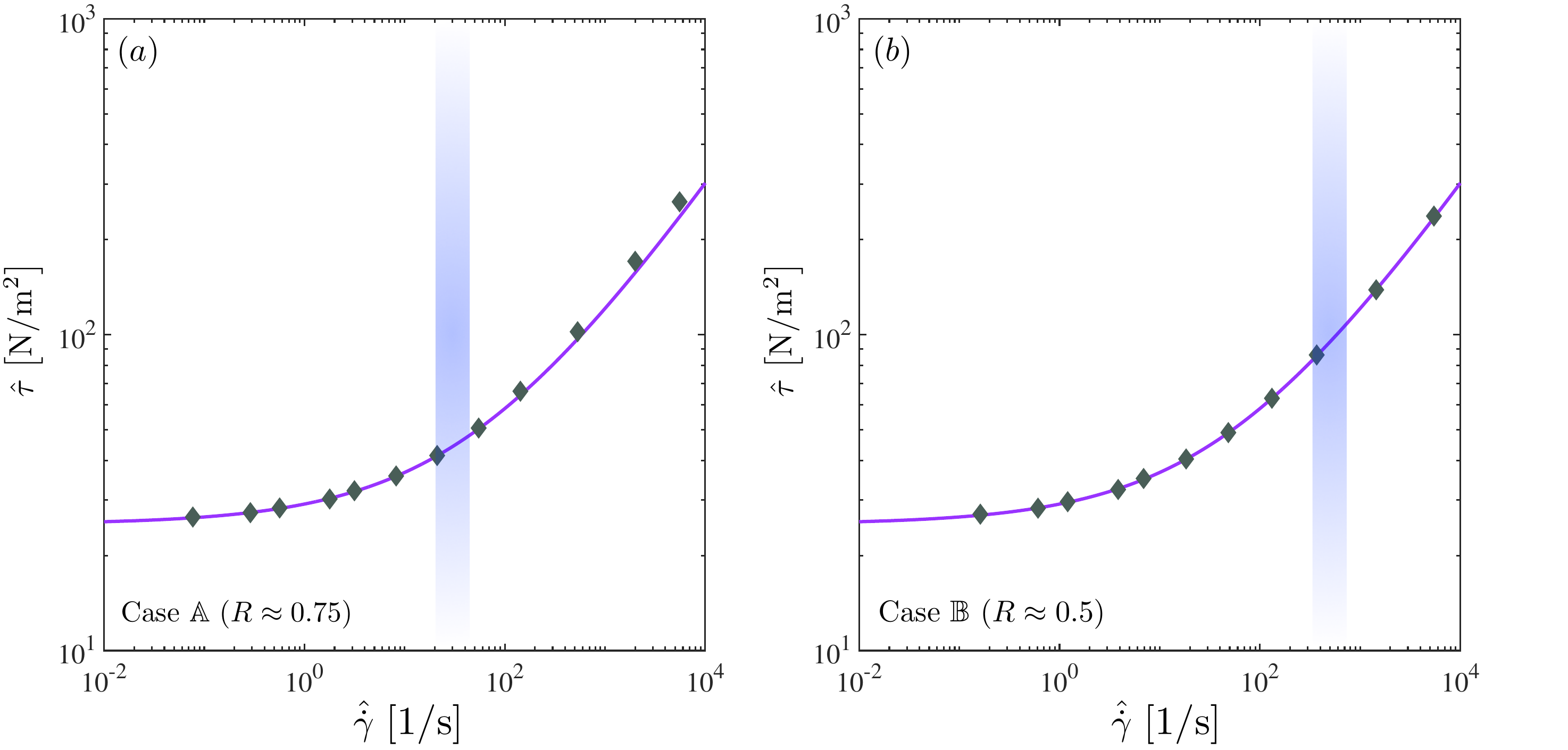}}
\caption{Flow curves of a Herschel-Bulkley fluid ($\hat{\tau}_y=25.1 ~\text{Pa}, \hat{K}=4 ~\text{Pa}\cdot \text{s}^\text{n} ~\&~ n=0.46$): the symbols are calculated based on the 1D circular Couette flow of Appendix \ref{sec:Couette} for (a) case $\mathbb{A}$ and (b) case $\mathbb{B}$ using expression (\ref{eq:wide_gap}). The purple lines are the actual flow curve. The blue boxes mark the transition between the fully yielded and partially yielded gaps in the circular Couette analogy.}
\label{fig:Couette_flowcurve}
\end{figure*}

As discussed in \S~\ref{sec:2D}, expression (\ref{eq:wide_gap}) was derived to convert the vane rotation rate into the corresponding nominal strain rate. In this appendix, we assess the accuracy of this conversion, particularly the validity of neglecting $f(\hat{\tau}_2)$ in expression (\ref{eq:gamma_full})), using 1D circular Couette flow solutions for both the narrow-gap (case $\mathbb{A}$) and wide-gap (case $\mathbb{B}$) configurations.

Fig.~\ref{fig:Couette_flowcurve} compares the flow curves calculated using this conversion for (a) the narrow-gap configuration $\mathbb{A}$ and (b) the wide-gap configuration $\mathbb{B}$. The solid purple lines represent the prescribed flow curves, while the diamonds denote the reconstructed flow curves obtained by applying expression (\ref{eq:wide_gap}) to the circular Couette solutions presented in Appendix \ref{sec:Couette}. The blue boxes indicate the transition between the fully yielded and partially yielded regimes. As shown, the calculated flow curves closely follow the prescribed one for both configurations, with slightly better agreement observed for the wide-gap case $\mathbb{B}$. Notably, even at high shear rates $\sim O(10^4)$ shown in the figure, the approximation $\vert f(\hat{\tau}_2)\vert \ll \vert f(\hat{\tau}_1)\vert$, or equivalently $\vert \hat{\dot{\gamma}}_2\vert \ll \vert \hat{\dot{\gamma}}_1\vert$, remains well justified, even for the narrow-gap configuration.

\section*{References}

\bibliography{Viscoplastic.bib}% Produces the bibliography via BibTeX.

\end{document}